\documentclass
[twocolumn,aps,prd,amsmath,amssymb,floatfix]
{revtex4-2}
\usepackage{silence} 
\usepackage{CJK}
\usepackage[dvips]{graphicx}
\usepackage{bm}                      
\usepackage{mathptmx}                
\usepackage{dcolumn}                 
\usepackage{multirow}                
\usepackage{xcolor}                  %
\graphicspath{{eps}}
\usepackage[
breaklinks,pdfstartview=FitH,CJKbookmarks=true,
bookmarksnumbered=true,bookmarksopen=true,pdfborder={0 0 1},
colorlinks=true,linkcolor=blue,urlcolor=blue,anchorcolor=blue,citecolor=blue]
{hyperref}

\begin{document}

\def\be{\begin{equation}} \def\ee{\end{equation}}
\def\bal#1\eal{\begin{align}#1\end{align}}
\def\ra{\rightarrow}
\def\arsinh{\mathop{\text{arsinh}}}
\def\al{\alpha}
\def\de{\Delta}
\def\eps{\epsilon}
\def\la{\Lambda}
\def\ms{\,M_\odot}
\def\mmax{M_\text{max}}
\def\fm3{\;\text{fm}^{-3}}
\def\gc3{\;\text{g/cm}^3}
\def\km{\;\text{km}}
\def\mev{\;\text{MeV}}
\def\gev{\;\text{GeV}}
\def\rdu{\rho_\text{DU}} \def\mdu{M_\text{DU}} \def\xdu{x_\text{DU}}
\def\r1s0{\rho_{1S0}} \def\m1s0{M_{1S0}}
\def\mcr{M_\text{cr}}
\def\md{\mu} \def\fd{f}

\long\def\hj#1{\color{red}#1\color{black}}
\long\def\OFF#1{}


\title{Cooling of dark neutron stars}

\begin{CJK*}{UTF8}{gbsn}

\author{B. X. Zhou (周博修)$^1$}
\author{H. C. Das$^2$}
\author{J. B. Wei (魏金标)$^3$}
\author{G. F. Burgio$^2$}
\author{Z. H. Li (李增花)$^{1,4}$}\email{zhli09@fudan.edu.cn}
\author{H.-J. Schulze$^2$}

\affiliation{
$^1$Institute of Modern Physics,
Key Laboratory of Nuclear Physics and Ion-Beam Application, MOE,
Fudan University, Shanghai 200433, China}
\affiliation{
$^2$INFN Sezione di Catania, Dipartimento di Fisica,
Universit\'a di Catania, Via Santa Sofia 64, 95123 Catania, Italy}
\affiliation{
$^3$Physics Department, University of Geosciences, Wuhan, P.R.~China}
\affiliation{
$^4$Shanghai Research Center for Theoretical Nuclear Physics, NSFC,
Fudan University, Shanghai, 200438, China
}

\date{\today}

\begin{abstract}
We study the cooling of isolated dark-matter-admixed neutron stars,
employing a realistic nuclear equation of state
and realistic nuclear pairing gaps,
together with fermionic dark matter
of variable particle mass and dark-matter fraction.
The related parameter space is scanned for the
stellar structural and cooling properties.
We find that a consistent description of all current cooling data
requires fast direct Urca cooling and reasonable proton 1S0 gaps.
Dark matter affects the cooling properties by a modification
of the nuclear density profiles,
but also changes stellar radius and maximum mass.
Possible signals of a large dark matter content
could be a very massive but slow-cooling star
or a very light but fast-cooling star.
\end{abstract}

\maketitle
\end{CJK*}


\section{Introduction}

The enigma of dark matter (DM)
\cite{Zwicky09,Trimble87,Bergstrom00,Bertone05,Feng10,Bertone18,Bramante24}
is currently causing frenetic theoretical activity,
trying to identify unequivocal signals for its presence and properties
in various observable natural environments.
In particular, neutron stars (NSs) have been identified as possible
candidates for such signals,
due to their enormous gravitational field and consequent potential capability
of capturing large amounts of DM,
which is otherwise sterile to the other three fundamental interactions.
In particular, possible effects on global observables of such dark NSs (DNSs)
like their mass, radius, tidal deformability, etc.~have been quantitatively
investigated in many recent publications, see
\cite{Gresham19,
Das20,Das20_curv,Das21_inspiral,Das21_fmode,Das21_GW19,Das21_Galaxy,
Sen21,Kumar22,Dutra22,Karkevandi22,Rutherford23,Mu23,Routaray23,Liu23,
Caballero24,Liu24,Mariani24,Routaray25,Kumar25}
for recent references.

A more delicate potential impact of DM is the one on the
cooling properties of DNSs,
observationally accessible in terms of
temperature (or luminosity) vs age relations,
and which are an important tool to obtain a glimpse
on the internal structure and composition of NS matter
\cite{Yakovlev01,Yakovlev14,Miller21b,Burgio21},
complementary to the global NS observables.

Recently, several works
\cite{Bhat20,Kumar22,Garcia22,Avila24,Giangrandi24}
have pointed out an interesting effect of DM
on the cooling due to a modification of the range of direct Urca (DU)
cooling inside the star:
Moderate amounts of DM will shrink the nuclear core of a DNS,
increase the central nuclear density,
and thus extend the fast-cooling domain
and cause stronger cooling in a certain DNS mass range.
However, apart from the fact that a sharp onset of DU cooling is probably
an unrealistic approximation \cite{Beznogov15,Burgio21},
current investigations of this kind are restricted
to very small DM fractions of a few percent at most,
and thus do hardly exhibit the principal effects of DM, due to
(a) reduction of the active nucleonic cooling mass/volume of the star,
(b) changed composition inside that volume
for a fixed total gravitational mass,
(c) most important, impact on superfluidity thresholds,
which in fact smoothens the DU onset in realistic models.
Apart from this,
a substantial DM content changes the maximum mass of a DNS
(in both senses)
and thus the accessible cooling configurations.

It is the purpose of this work to perform such a study
over the full range of DM fraction $f=M_D/M$
(ratio of dark and total gravitational mass of the DNS)
from 0 (pure NS) to 1 (pure dark star),
covering DM-core and DM-halo stars.
We will use a combination of a realistic nucleonic equation of state (EOS)
with a standard fermionic asymmetric DM model
characterized effectively by one parameter $\md$, the DM particle mass.
Therefore the DM degrees of freedom are $\md$ and $f$,
and their effects will be studied in detail.
As this kind of DM is not interacting directly with visible matter,
all effects on the cooling are indirect,
by the action of the modified total gravitational field
on the normal matter.
We emphasize already that such a scenario of very large DM fractions
is very speculative and requires exotic formation mechanisms of DNSs like
dark conversion of the neutron to scalar DM,
neutron bremsstrahlung of scalar DM,
mirror DM,
self-interaction of condensate DM
\cite{Foot04,Sandin09,XYLi12,Perez22,Garcia22},
since the standard DM capture mechanism of diluted DM in the Universe
\cite{Ellis18}
is only able to provide DM fractions of about $f\sim10^{-10}$
for NSs in the galactic centers.
Therefore even DM fractions of a few percent
that have been assumed in many articles,
are as extreme as 10 or 90 percent,
and demand nonstandard explanations.
With this in mind we perform this study as continuation of \cite{Liu23,Liu24},
where we analyzed the global properties of $f=0,...,1$ DNSs.

NS cooling data are now available for about 60 objects
\cite{Potekhin20,Cooldat},
which has encouraged first tentative combined analyses of cooling properties
and NS mass distributions,
as initiated in
\cite{Popov06,Blaschke07,Beznogov15,Beznogov15b,Wei19,Wei20,Das24}.
In particular,
as investigated in our previous articles on this topic
\cite{Taranto16,Fortin18,Wei19,Wei20,Wei20b,Das24},
it has become increasingly clear
\cite{Beznogov15,Beznogov15b,Potekhin20,Leinson22,Burgio21}
that fast neutrino cooling processes
like the DU reaction
are required in order to explain
cold and not too old objects,
while also sufficiently slow cooling must be accommodated theoretically
to cover old and warm objects.
These data will also allow us to interpret the predictions for DNSs
obtained here.
It is therefore essential to base the theoretical analysis on a
framework that is able to explain {\em all} current cooling data
in a reasonable manner,
and not fine-tune the analysis to certain objects or NS masses,
at the cost of disregarding the rest of the data available.

Regarding the nuclear EOS,
recent observations of massive NSs
\cite{Antoniades13,Arzoumanian18,Cromartie20,Fonseca21,Romani22,
Riley19,Miller19,Riley21,Miller21,Pang21,Raaijmakers21,
Vinciguerra24,Salmi24,Dittmann24,Choudhury24,Salmi24b}
and gravitational-wave observations in particular
\cite{Abbott18,Radice18,Raaijmakers21,Burgio21}
have significantly restricted the possible choice
of a theoretical EOS \cite{Wei20c,Burgio21}.
We employ here an EOS based on the Brueckner-Hartree-Fock (BHF) approach
that fulfills all presently known constraints of this kind.
This is very important, as practically all currently observed NSs (pulsars)
are supposed to be pure nucleonic stars,
and DNSs with a large DM fraction as discussed in this work
are thought to be very rare and exotic objects,
if they exist at all.
To indicate their particular properties is the purpose of this work.

This paper is organized as follows.
In Sec.~\ref{s:eos} we give a brief overview of the theoretical framework,
regarding the nuclear and DM EOSs,
the various cooling processes,
the related nucleonic pairing gaps,
and the cooling simulations.
In Sec.~\ref{s:resa} we present results regarding the EOS
and the structure of DNSs.
Sec.~\ref{s:resb} is devoted to the presentation and discussion
of the cooling diagrams and their dependence on the various
theoretical degrees of freedom.
Conclusions are drawn in Sec.~\ref{s:end}.
We use natural units $G=c=\hbar=1$ throughout the article.

\section{Formalism}
\label{s:eos}

In this work we assume purely nucleonic baryonic matter, i.e.,
pure NS matter is composed of nucleons and leptons only.
Exotic components like hyperons or quark matter are not considered
\cite{Burgio21}.

\subsection{Equation of state for nuclear matter}

As in \cite{Liu23,Liu24,Das24},
we employ here the latest version of a BHF EOS
obtained with the Argonne V18 nucleon-nucleon potential
and compatible three-body forces
\cite{Li08a,Li08b,Lu19,Liu22},
see \cite{Baldo99,Baldo12} for a more detailed account.
This EOS is compatible with all current low-density constraints
\cite{Wei20,Burgio21,Burgio21b}
and in particular also with those imposed on NS maximum mass $\mmax>2\ms$
\cite{Antoniades13,Arzoumanian18,Cromartie20,Fonseca21},
including the recent $M=2.35\pm0.17\ms$ for PSR J0952-0607 \cite{Romani22},
the mass-radius results of the NICER mission for the pulsars
J0030+0451 \cite{Riley19,Miller19,Vinciguerra24},
J0740+6620 \cite{Riley21,Miller21,Pang21,Raaijmakers21,Salmi24,Dittmann24}, and
J0437-4715 \cite{Choudhury24},
as well as the tidal deformability range $\Lambda_{1.4}\approx70-580$
of GW170817 \cite{Abbott17,Abbott18,Burgio18,Wei19}.

We remind that the BHF EOS describes homogeneous nuclear matter,
and therefore an EOS for the low-density inhomogeneous crustal part
has to be added.
For that,
we adopt the well-known Negele-Vautherin EOS \cite{Negele73}
for the inner crust in the medium-density regime
($0.001\fm3 < \rho < \rho_t$),
and the ones by Baym-Pethick-Sutherland \cite{Baym71} and
Feynman-Metropolis-Teller \cite{Feynman49} for the outer crust
($\rho < 0.001\fm3$).
By imposing a smooth transition
of pressure and energy density between both branches of the betastable EOS
\cite{Burgio10},
one finds a transition density at about $\rho_t \approx 0.08\fm3$.
In any case the NS maximum-mass domain is not affected by the crustal EOS,
with a limited influence on the radius and related deformability
for NSs with canonical mass value
\cite{Burgio10,Baldo14,Fortin16,Tsang19}.

\subsection{Equation of state for dark matter}

We employ in this work the frequently-used
\cite{Narain06,Li12,Tulin13,Kouvaris15,Maselli17,Ellis18,Nelson19,
Delpopolo20b,Husain21,Leung22,Collier22,Cassing23,
Liu24,Avila24,Giangrandi24,Routaray25}
DM model of fermions with mass $\md$,
self-interacting via a repulsive Yukawa potential
\bal
 V(r) &= \al \frac{e^{-m r}}{r} \:
\eal
with coupling constant $\al$ and mediator mass $m$.
Following Ref.~\cite{Narain06} we write
pressure and energy density
of the resulting DM EOS as
\bal
 p_D    &= \frac{\md^4}{8\pi^2}
 \Big[ x\sqrt{1+x^2}(2x^2\!/3-1) + \arsinh{(x)} \Big]
 + \delta \:,
\\
 \eps_D &= \frac{\md^4}{8\pi^2}
 \Big[ x\sqrt{1+x^2}(2x^2+1) - \arsinh{(x)} \Big]
 + \delta \:,
\eal
where
\be
 x = \frac{k_F}{\md} = \frac{(3\pi^2n)^{1/3}}{\md}
\ee
is the dimensionless kinetic parameter
with the DM particle density $n$,
and the self-interaction term is written as
\bal
 \delta =
 \frac{2}{9\pi^3} \frac{\al \md^6}{m^2} x^6 &\equiv
  \md^4 \Big(\frac{y}{3\pi^2}\Big)^2 x^6
 = \Big(\frac{yn}{\md}\Big)^2 \:,
\label{e:y}
\eal
introducing the interaction parameter
$y^2 = 2\pi\al\md^2\!/m^2$.
Ref.~\cite{Narain06} contains interesting scaling relations
regarding the EOS and mass-radius relations of pure fermionic DM stars.

Within this model,
$\md$ and $y$ are not independent free parameters,
but constrained
by limits imposed on the DM self-interaction cross section $\sigma$
through observation of the interaction of galaxies
in different colliding galaxy clusters
\cite{Markevitch04,Kaplinghat16,Sagunski21,Loeb22},
\be
 \sigma/\md \sim 0.1-10 \;\text{cm}^2\!/\text{g} \:.
\ee
In \cite{Tulin13,Kouvaris15,Maselli17}
it has been shown that the Born approximation
\bal
 \sigma_\text{Born} &= \frac{4\pi\al^2}{m^4}\md^2 = \frac{y^4}{\pi\md^2}
\eal
is very accurate for $\md\lesssim1\gev$
and in any case remains valid in the limit $\al\ra0$ for larger masses.
We therefore employ here this approximation,
choosing for simplicity the fixed constraint
\be
 \sigma/\md = 1 \;\text{cm}^2\!/\text{g} = 4560/\gev^3 \:,
\ee
which appears compatible with all current observations.
This implies
\bal
 y^4 &= \pi \md^3 \sigma/\md \sim \pi (16.58\md_1)^3
\:,\\
 y &\sim 10.94 \md_1^{3/4}
\label{e:ymu}
\eal
with $\md_1\equiv\md/1\!\gev$.
After this, the DM EOS depends only on the single parameter $\md$.

\OFF{
This is demonstrated in Fig.~\ref{f:eos},
which compares the DM EOS with and without self interaction for four
typical masses $\md=0.5,1,2,4 \gev$,
in comparison with the nucleonic EOS.
Markers indicate the configurations of maximum stellar mass.
It can be seen that the self-interaction is always important in the
relevant density domains
and stiffens substantially the EOS.
The nucleonic EOS exhibits
different trends for core, inner crust, and outer crust of a NS.
}

\subsection{Hydrostatic configuration and stellar structure}

The stable configurations of DNSs are obtained from a
two-fluid version of the TOV equations
\cite{Kodama72,Comer99,Sandin09}:
\bal
 \frac{dp_D}{dr} &= -(p_D + \eps_D)\frac{d\phi}{dr} \:,
\label{e:tovd}
\\
 \frac{dp_N}{dr} &= -(p_N + \eps_N)\frac{d\phi}{dr} \:,
\label{e:tovn}
\\
 \frac{dm}{dr}   &= 4\pi r^2 \eps \:,
\label{e:tovm}
\\
 \frac{dN_i}{dr} &=
 4\pi r^2 n_i e^{\lambda} \:,
\label{e:tovni}
\\
 \frac{d\phi}{dr} &=
 \frac{m + 4\pi r^3p}{r^2}  e^{2\lambda}\:,
\label{e:tovnu}
\eal
where $r$ is the radial coordinate from the center of the star,
$\phi$ and $e^{\lambda}=1/\sqrt{1-2m/r}$ are metric functions,
and
$p=p_N+p_D$,
$\eps=\eps_N+\eps_D$,
$m=m_N+m_D$
are the total pressure, energy density, and enclosed mass, respectively.
In addition, $N_i$ are the enclosed particle numbers
with the corresponding densities $n=k_F^3/3\pi^2$.

The total gravitational mass of the DNS is
\be
 M = m_N(R_N) + m_D(R_D) \:,
 \label{eq:gravmass}
\ee
where the stellar radii $R_N$ and $R_D$
are defined by the vanishing of the respective pressures.
There are thus in general two scenarios:
DM-core ($R_D<R_N$) or DM-halo ($R_D>R_N$) stars.

It has been shown in stability analyses of the two-fluid model
\cite{Henriques90b,Leung12,Valdez13,Kain20,Kain21,Leung22,Gleason22,
Hippert23,Barbat24,Zhen24,Pitz25,Biesdorf25}
that the limits of stellar stability occur at the
maximum of $M$ along fixed-particle-number contours,
and this has been demonstrated \cite{Leung12,Kain20,Leung22,Zhen24}
to coincide quite accurately with the maximum of the $M(R)$
relations for fixed DM fraction $f$,
just as in the limit of ordinary NSs.
Configurations with larger central densities (smaller radius)
are unstable with respect to induced radial oscillations,
as verified in the above references.
We show and discuss only stable DNS configurations in the following,
just as in our previous articles \cite{Liu23,Liu24}.

The nuclear EOS $p_N(\eps_N)$ in Eq.~(\ref{e:tovn})
is the one of betastable and charge-neutral matter,
as for ordinary NSs.
It is important to note that due to the absence
of strong and weak interactions involving DM, this EOS
(and the related particle fractions as functions of nucleon density)
are unaffected by the presence of DM.
This strongly simplifies also the cooling simulations.

\subsection{Cooling processes}

In the context of NS cooling the one key property of the nuclear EOS
is whether it allows fast direct Urca (DU) cooling
by a large enough proton fraction
\cite{Yakovlev01,Page06,Page06b,Lattimer07,Yakovlev14,Potekhin15,Burgio21}.
The DU process is by 
several orders of magnitude
the most efficient one among all possible
cooling reactions involving nucleons and neutrino emission
and depends on the NS EOS and composition.
It involves a pair of charged weak-current reactions,
\be
 n \ra p + l + \bar{\nu}_l
\quad \text{and} \quad
 p + l \ra n + \nu_l \:,
\label{e:DU}
\ee
being $l=e,\mu$ a lepton and $\nu_l$ the corresponding neutrino.
Those reactions are allowed by energy and momentum conservation
only if $k_F^{(n)} < k_F^{(p)} + k_F^{(l)}$,
where $k_F^{(i)}$ is the Fermi momentum of the species $i$
\cite{Lattimer91}.
This implies that the proton fraction should be larger than a threshold value
(about 13\%, depending on the muon fraction)
for the DU process to take place,
and therefore the NS central density should be larger
than the corresponding nuclear threshold density $\rdu$.

Besides the DU process,
other cooling reactions come into play and involve nucleon collisions,
the strongest one being the modified Urca (MU) process,
\be
 n + N \ra p + N + l \! + \bar{\nu}_l
\quad \text{and} \quad
 p + N + l \ra n + N + \nu_l \:,
\label{e:MU}
\ee
where $N$ is a spectator nucleon that ensures momentum conservation.
The nucleon-nucleon bremsstrahlung (BS) reactions,
\be
 N+N \ra N+N + \nu+\bar{\nu} \:,
\ee
with $N$ a nucleon and $\nu$,
$\bar{\nu}$ an (anti)neutrino of any flavor,
are also abundant in NS cores,
and their rate increases with the nucleon density,
but they are orders of magnitude less powerful than the DU one,
thus producing a slow cooling \cite{Yakovlev01}.
All those cooling mechanisms can be strongly affected
by the superfluid properties of the stellar matter, i.e.,
critical temperatures and gaps in the different pairing channels
\cite{Yakovlev01,Sedrakian19,Burgio21}.

\subsection{Pairing gaps}
\label{s:gap}


Usually the most important ones are the proton 1S0 (p1S0)
and neutron 3PF2 (n3P2) pairing channels;
the proton 3PF2 gap is often neglected due to its uncertain properties
at large densities,
while the neutron 1S0 gap in the crust is of little relevance for the cooling.

These superfluids are created by the formation of $pp$ and $nn$ Cooper pairs
due to the attractive part of the $NN$ potential,
and are characterized by a critical temperature
$T_c \approx 0.567\Delta$.
The occurrence of pairing when $T < T_c$
leads on one hand to an exponential reduction of the emissivity
of the neutrino processes the paired component is involved in,
and on the other hand to the onset of the ``pair breaking and formation"
(PBF) process with associated neutrino-antineutrino pair emission.
This process starts when the temperature reaches $T_c$
of a given type of nucleons,
becomes maximally efficient when $T \approx 0.8\,T_c$,
and then is exponentially suppressed for $T \ll T_c$ \cite{Yakovlev01}.

\OFF{
In the simplest BCS approximation,
the relevant pairing gaps
are computed by solving the (angle-averaged) gap equation
\cite{Amundsen85,Baldo92,Takatsuka93,Elgaroy96,Khodel98,Baldo98}
for the $L=0$ and the two-component $L=1,3$ gap functions,
\bal
 \de_0(k) &= - \frac{1}{\pi} \int_0^{\infty}\!\! dk' {k'}^2 \frac{1}{E(k')}
  V_{00}(k,k') \de_0(k') \:,
\label{e:gaps}
\\
 \left(\!\!\!\begin{array}{l} \de_1 \\ \de_3 \end{array}\!\!\!\right)\!(k) &=
 - \frac{1}{\pi} \int_0^{\infty}\!\! dk' {k'}^2 \frac{1}{E(k')}
 \left(\!\!\!\begin{array}{ll}
  V_{11}\!\! & \!\!V_{13} \\ V_{31}\!\! & \!\!V_{33}
 \end{array}\!\!\!\right)\!(k,k')
 \left(\!\!\!\begin{array}{l} \de_1 \\ \de_3 \end{array}\!\!\!\right)\!(k')
\label{e:gapp}
\eal
with
\bal
 E(k)^2 &= [e(k)-\mu]^2 +
 \de(k)^2 \:, \\ \de(k)^2 &=
 \{ \de_0(k)^2 , \de_1(k)^2 + \de_3(k)^2 \} \:,
\eal
while fixing the (neutron or proton) density,
\be
  \rho = \frac{k_F^3}{3\pi^2}
   = 2 \sum_k \frac{1}{2} \left[ 1 - \frac{e(k)-\mu}{E(k)} \right] \:.
\label{e:rho}
\ee
Here $e(k)=k^2\!/2m$ is the s.p.~energy,
$\mu \approx e(k_F)$ is the chemical potential
determined self-consistently from Eqs.~(\ref{e:gaps}--\ref{e:rho}),
and
\be
   V^{}_{LL'}(k,k') =
   \int_0^\infty \! dr\, r^2\, j_{L'}(k'\!r)\, V^{TS}_{LL'}(r)\, j_L(kr)
\label{e:v}
\ee
are the relevant potential matrix elements
($T=1$ and
$S=1$; $L,L'=1,3$ for the 3PF2 channel,
$S=0$; $L,L'=0$ for the 1S0 channel)
with the bare potential $V$. 
The relation between (angle-averaged) pairing gap at zero temperature
$\de \equiv \de(k_F)$
obtained in this way and the critical temperature of superfluidity is then
$T_c \approx 0.567\de$.
}

As in previous work \cite{Wei20,Das24},
to which we refer for more details,
we use here a scaled p1S0 BCS gap,
\be
 \Delta(\rho) \equiv s_y \Delta_\text{BCS}(\rho/s_x) \:,
\label{e:sxy}
\ee
to simulate medium effects beyond the BCS approximation,
with optimal scaling factors $s_x=0.8,s_y=0.6$ \cite{Wei20,Das24},
and exclude a n3P2 gap,
since it conflicts with current observational data in our cooling simulations
\cite{Wei20,Wei20b}.

\begin{figure}[t]
\vspace{-3mm}
\centerline{\includegraphics[scale=0.72]{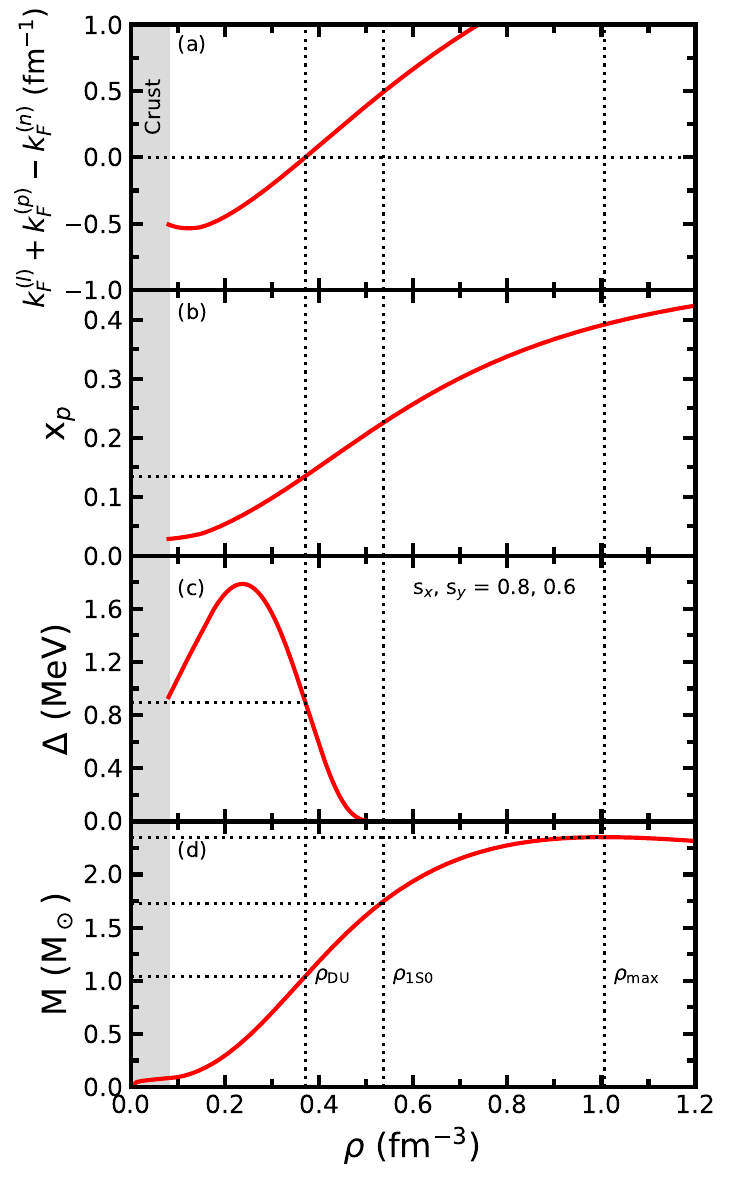}}
\vspace{-4mm}
\caption{
The threshold condition for the DU process (a),
the proton fraction (b),
the scaled p1S0 gap (c), and
the NS mass (d)
vs.~the nucleon (central) density for the V18 EOS.
Vertical lines indicate
the onset of DU cooling,
the disappearance of the p1S0 gap, and
the $\mmax$ configuration.
\hj{}
}
\label{f:xp}
\end{figure}

\begin{figure}[t]
\vskip-2mm
\centerline{\hskip0mm\includegraphics[scale=0.51]{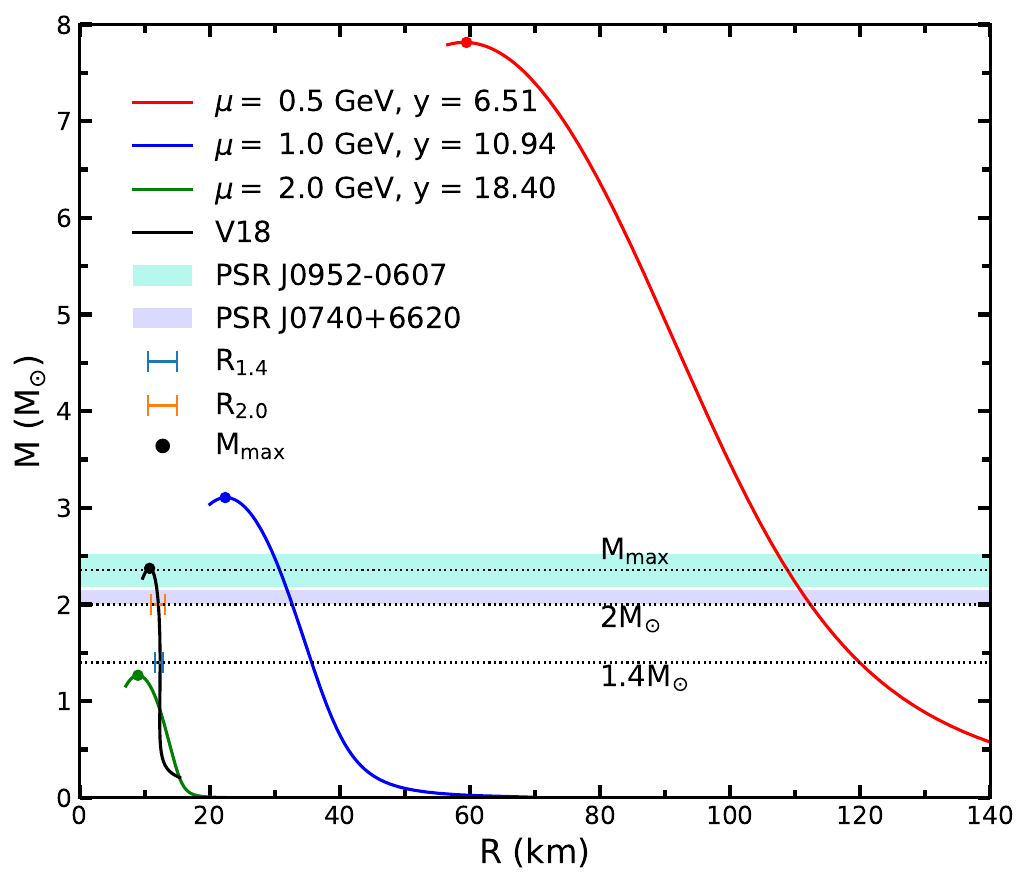}}
\vskip-3mm
\caption{
Mass-radius relations of pure dark stars
for different values of the DM particle mass $\md=0.5,1,2\gev$
[and associated interaction parameter $y$, Eq.~(\ref{e:ymu})]
(colored curves),
in comparison with the NM V18 EOS (black curve).
The horizontal lines indicate the maximum mass of a pure NS,
$\mmax=2.34\ms$, and $M=1.4,2.0\ms$.
Observational constraints on masses \cite{Cromartie20,Romani22}
and radii $R_{1.4}$ and $R_{2.0}$ from NICER \cite{Miller21,Rutherford24}
are included.
\hj{}
}
\label{f:mrd}
\end{figure}

\subsection{Cooling simulations}
\label{s:res}

For the NS cooling simulations we employ the widely known
one-dimensional code {NSCool} \cite{Pageweb},
based on an implicit scheme developed in \cite{Henyey64} for solving
the general-relativistic partial differential equations
of energy balance and energy transport,
\bal
 & \frac{e^{-\lambda-2\phi}}{4\pi r^2}
 \frac{\partial}{\partial r}(e^{2\phi}L_\gamma) + Q_\nu =
  - \frac{c_v}{e^\phi} \frac{\partial T}{\partial t} \:,
\\
 & \frac{L_\gamma}{4\pi r^2} =
 -\kappa e^{-\lambda-\phi}
 \frac{\partial}{\partial r}(e^\phi T) \:.
\eal
Local luminosity $L_\gamma$ and temperature $T$ depend on the
radial coordinate $r$ and time $t$.
The metric functions are $\phi$ and $\lambda$,
as in Eqs.~(\ref{e:tovd}-\ref{e:tovnu}).
$Q_\nu$, $c_v$, and $\kappa$ are the neutrino emissivity,
the specific heat capacity, and the thermal conductivity, respectively.
Observable quantities are the
surface temperature $T_s^\infty=T_s e^{-\lambda(R)}$
and photon luminosity $L_\gamma^\infty=L_\gamma e^{-2\lambda(R)}$.

The code solves the partial differential equations
on a grid of spherical shells.
The simulation is performed by artificially dividing
the star into two parts with a boundary radius $r_b$
at the nuclear density $\rho_b=10^{10}\gc3$.
At $\rho_N<\rho_b$,
the so-called envelope \cite{Beznogov21}
includes the mass and composition change,
for instance, due to the accretion,
and is treated separately in the code.
Here, we use the envelope model developed in \cite{Potekhin97}.
At $\rho_N>\rho_b\ (r<r_b)$,
the matter is strongly degenerate
and thus the structure of the star is supposed to be unchanged with time.
As a result,
for a NS of given mass and composition of the envelope,
we obtain a set of cooling curves showing the luminosity $L_\gamma^\infty$
as a function of NS age $t$.
Each simulation starts with an initial constant-temperature profile,
$\tilde{T}\equiv Te^{\phi}=10^{10}\,$K,
and ends when $\tilde{T}$ drops to $10^4\,$K.
Regarding the most important ingredient -- neutrino emissivity $Q_\nu$,
this code comprises all relevant cooling reactions:
nucleonic DU, MU, BS, and PBF,
including modifications due to p1S0 and n3P2 pairing.
The PBF rates are updated \cite{Leinson10}.
Moreover, various processes in the crust are included,
such as the most important electron-nucleus bremsstrahlung,
plasmon decay, electron-ion bremsstrahlung, etc.

The procedure is straightforwardly extended for DM-core stars.
However, care must be taken for DM-halo stars,
where the total energy density and pressure do not vanish
at the nucleonic visible surface of the star,
which might create numerical problems.
The nuclear crust and envelope must be treated accurately
in any configuration.

\begin{figure*}[t]
\vskip-13mm
\centerline{\hskip0mm\includegraphics[scale=0.6]{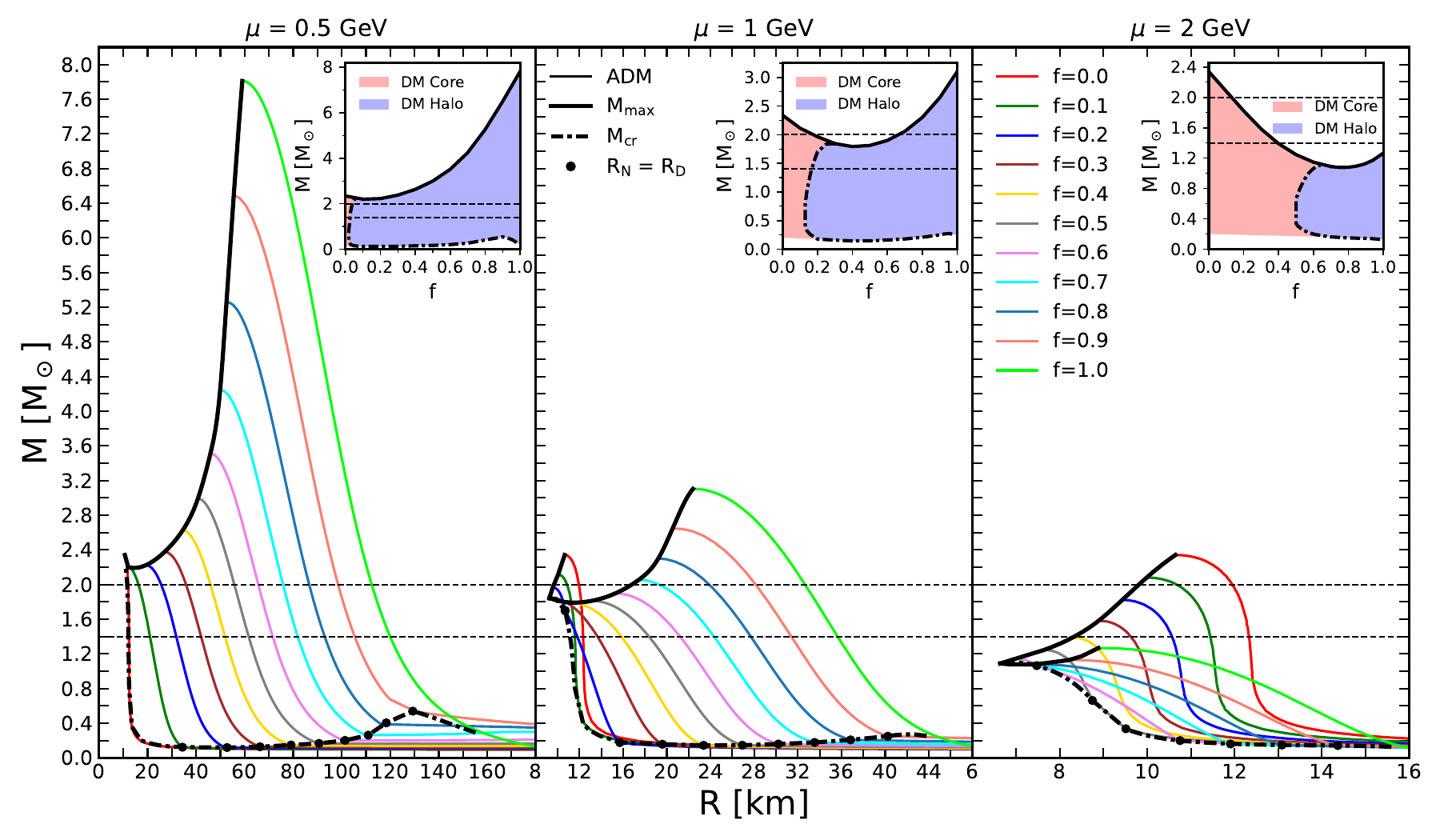}} 
\vskip-5mm
\caption{
The DNS mass-radius relations for fixed DM fractions $\fd=0,0.1,...,1$
(colored curves)
and for different DM particle masses $\md=0.5,1,2\gev$ (panels).
The solid black curves indicate the sequence of maximum masses.
The broken black curves connect the $R_N=R_D$ positions $M_\text{cr}$
on the fixed-$\fd$ curves.
The insets visualize both curves in the $(M,\fd)$ plane.
DM-core and DM-halo domains are indicated.
Note the different $R$ scales.
The $M/\ms=1.4,2.0$ horizontal lines are to guide the eye.
}
\label{f:mrf}
\end{figure*}

\begin{figure*}[t]
\vskip-3mm
\centerline{\hskip0mm\includegraphics[scale=0.6]{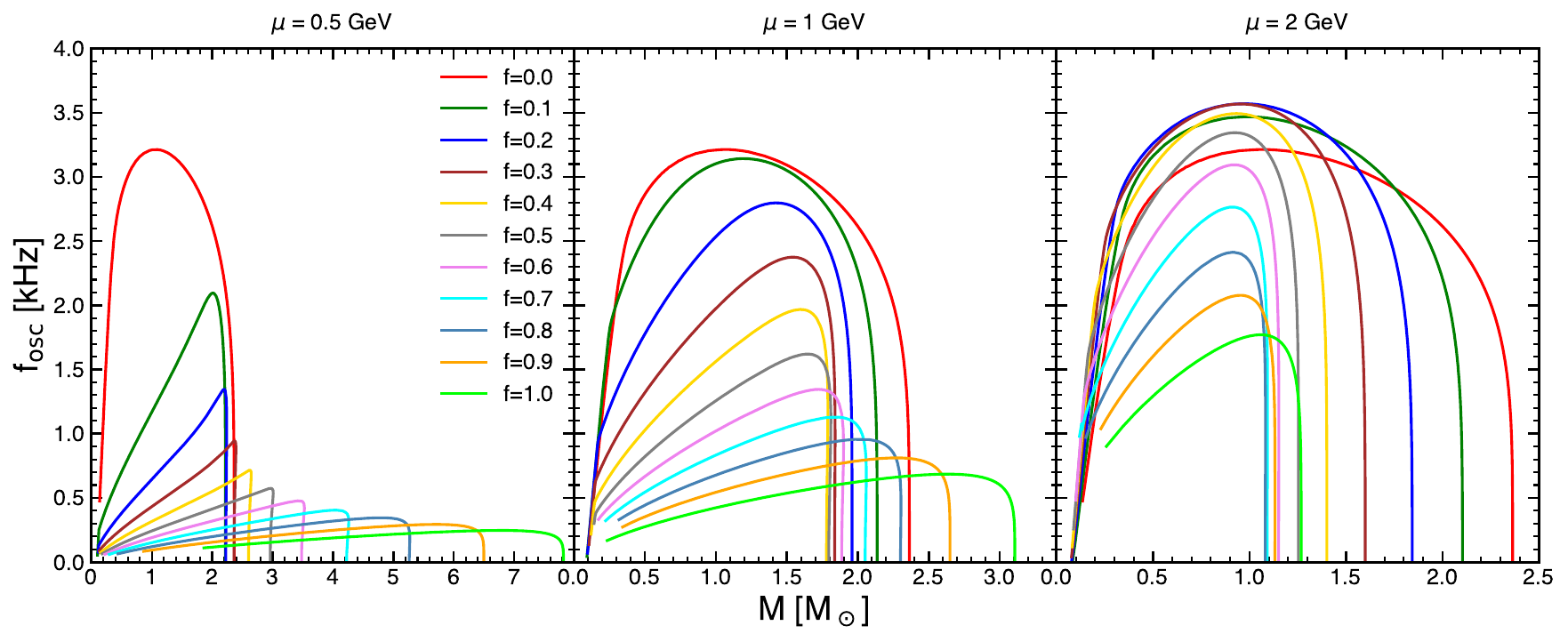}}
\vskip-5mm
\caption{
The radial oscillation frequencies of the DNS configurations in Fig.~\ref{f:mrf}.
}
\label{f:osc}
\end{figure*}

\OFF{
\begin{figure}[t]
\vskip-3mm
\centerline{\hskip0mm\includegraphics[scale=0.6]{mu_f1}}
\vskip-5mm
\caption{
Contours of the DM core-halo transition mass $\mcr$
in the $(\md,f)$ plane.
}
\label{f:mr}
\end{figure}
}

\begin{figure*}[t]
\vskip-2mm
\centerline{\hskip0mm\includegraphics[scale=0.55]{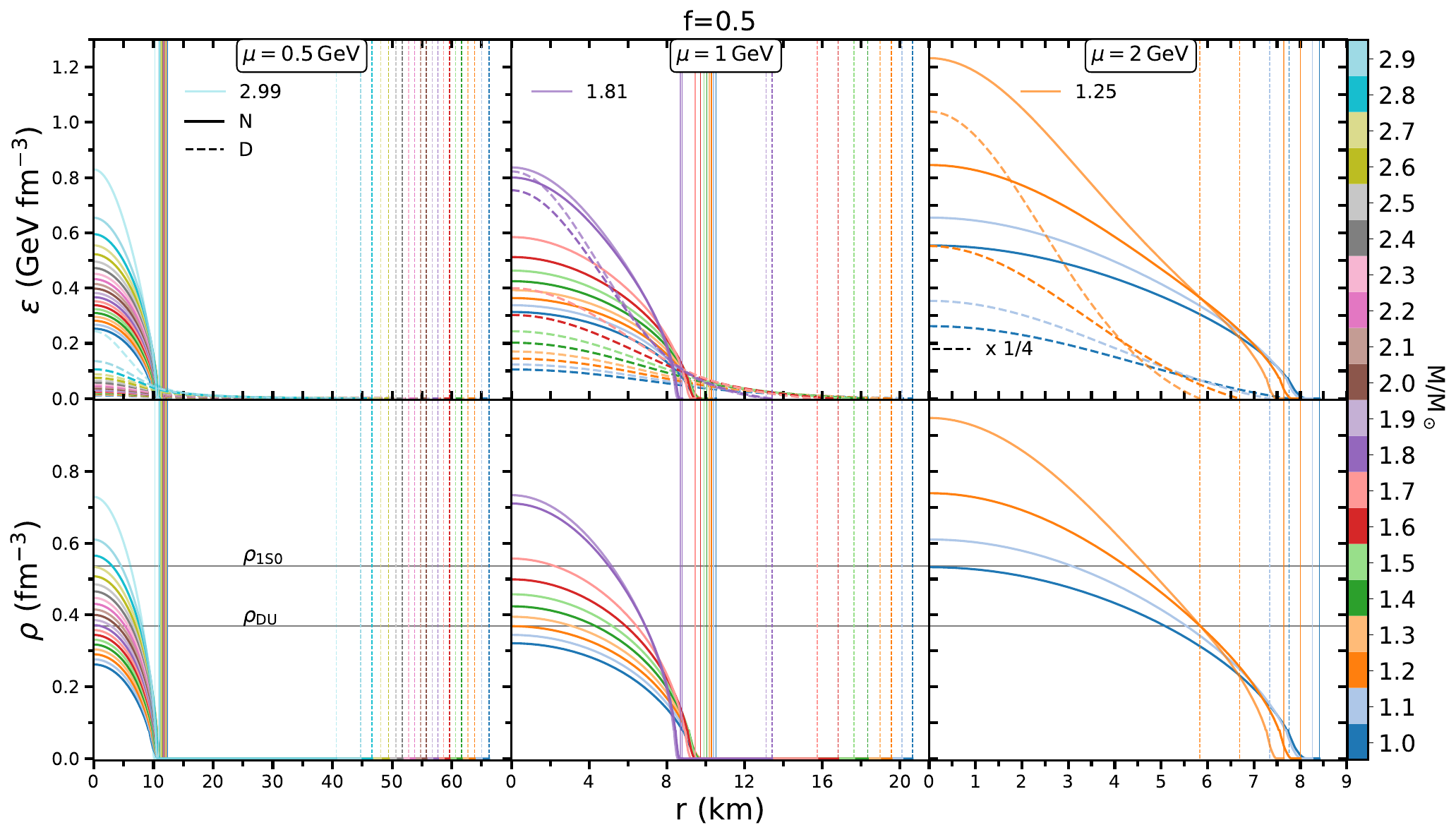}}
\vskip-4mm
\caption{
Profiles of
energy densities $\eps_N$ and $\eps_D$ (upper row)
and nucleon density $\rho$ (lower row)
for fixed $f=0.5$,
$\md=0.5,1,2\gev$,
$M=1.0\ms,1.1\ms,...,\mmax$ (indicated in the legend).
The vertical lines indicate the radii $R_N$ and $R_D$,
and the horizontal lines the densities of DU onset
and vanishing p1S0 gap.
In the top right panel, $\eps_D$ is scaled down by a factor 1/4.
Note the different $r$ scales.
\hj{}
}
\label{f:nrm}
\end{figure*}

\begin{figure*}[t]
\vskip-2mm
\centerline{\hskip0mm\includegraphics[scale=0.55]{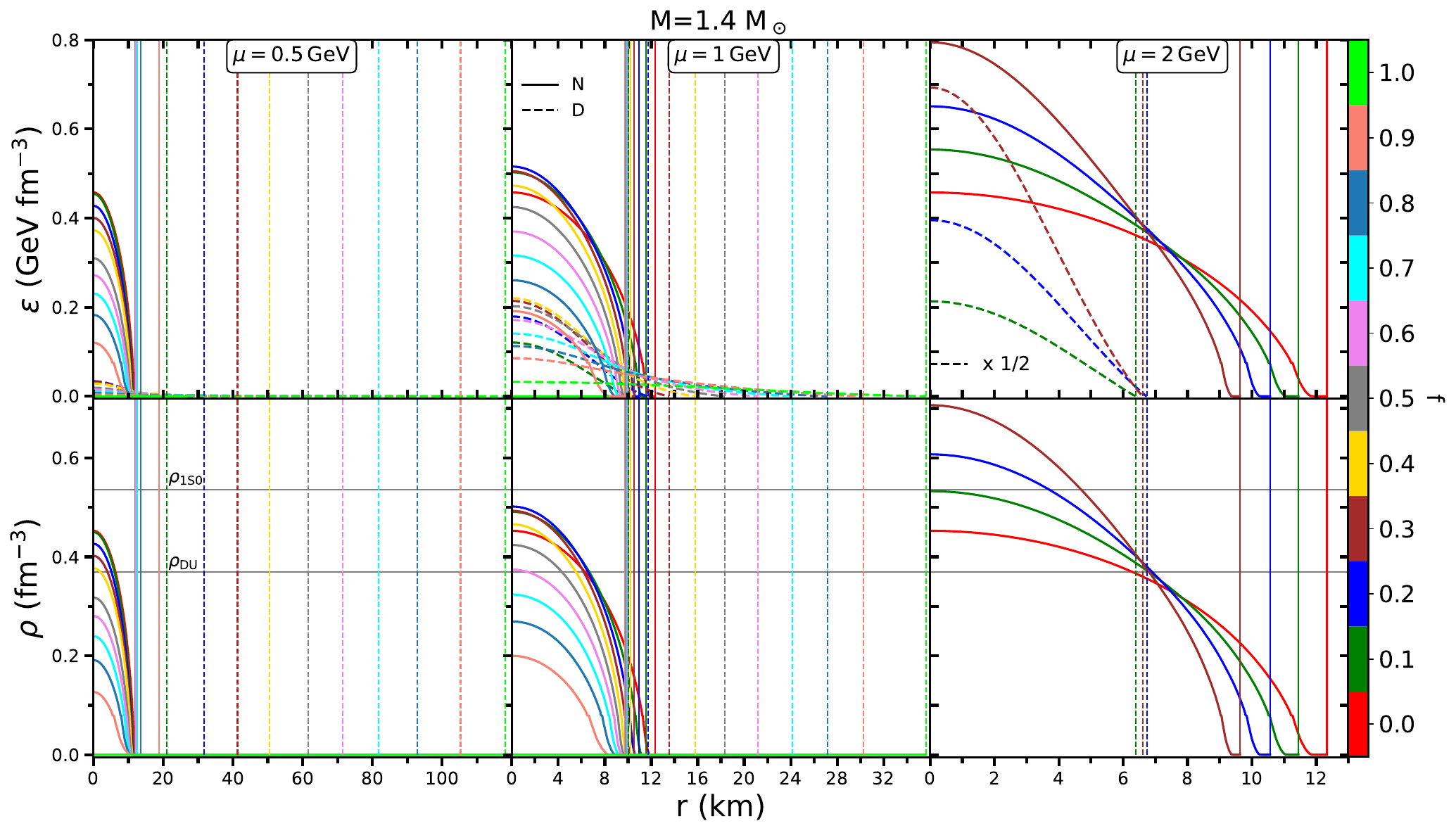}}
\vskip-4mm
\caption{
Same as Fig.~\ref{f:nrm},
but for fixed $M=1.4\ms$ and varying $f=0,0.1,\ldots,1$.
}
\label{f:nrf}
\end{figure*}

\OFF{
\begin{figure}[t]
\vskip-2mm
\centerline{\hskip0mm\includegraphics[scale=0.5]{r_nn_np}}
\vskip-4mm
\caption{
Density profiles
}
\label{f:mr}
\end{figure}
}

\begin{figure*}[t]
\vskip-2mm
\centerline{\hskip0mm\includegraphics[scale=0.5]{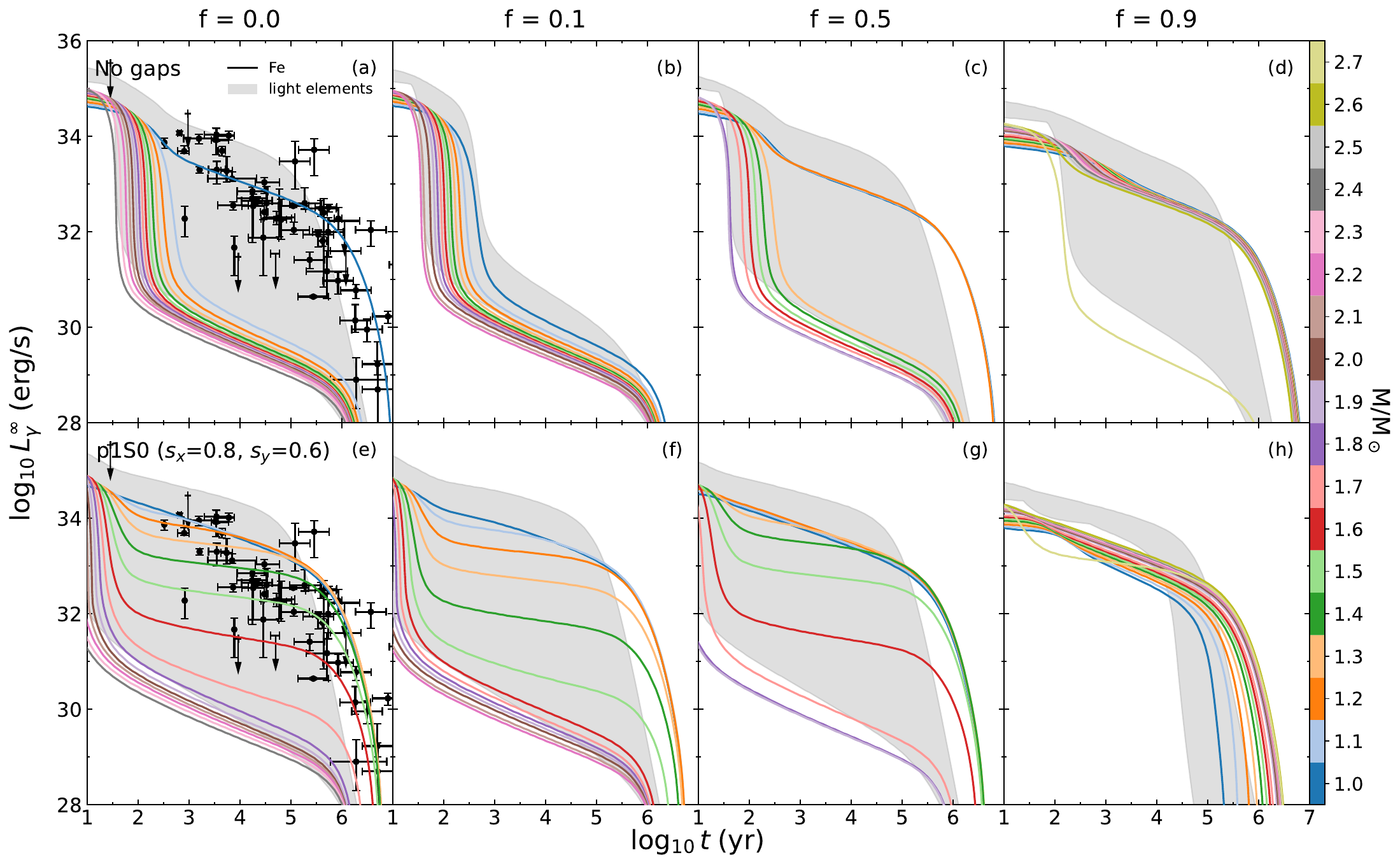}}
\vskip-3mm
\caption{
Cooling curves for DNSs with $\md=1\gev$ and
different DM fractions $f=0,0.1,0.5,0.9$ (columns)
without (upper row) and with (lower row) p1S0 gap
for different gravitational masses $M=1.0\ms,1.1\ms,\ldots,\mmax(f)$.
The curves are obtained with a Fe atmosphere,
while the ensembles of light-elements cooling curves are represented by the
shaded background areas.
The data points are from \protect\cite{Potekhin20,Cooldat}.
See text for details.
\hj{}
}
\label{f:coolm}
\end{figure*}

\begin{figure*}[t]
\vskip-2mm
\centerline{\hskip0mm\includegraphics[scale=0.5]{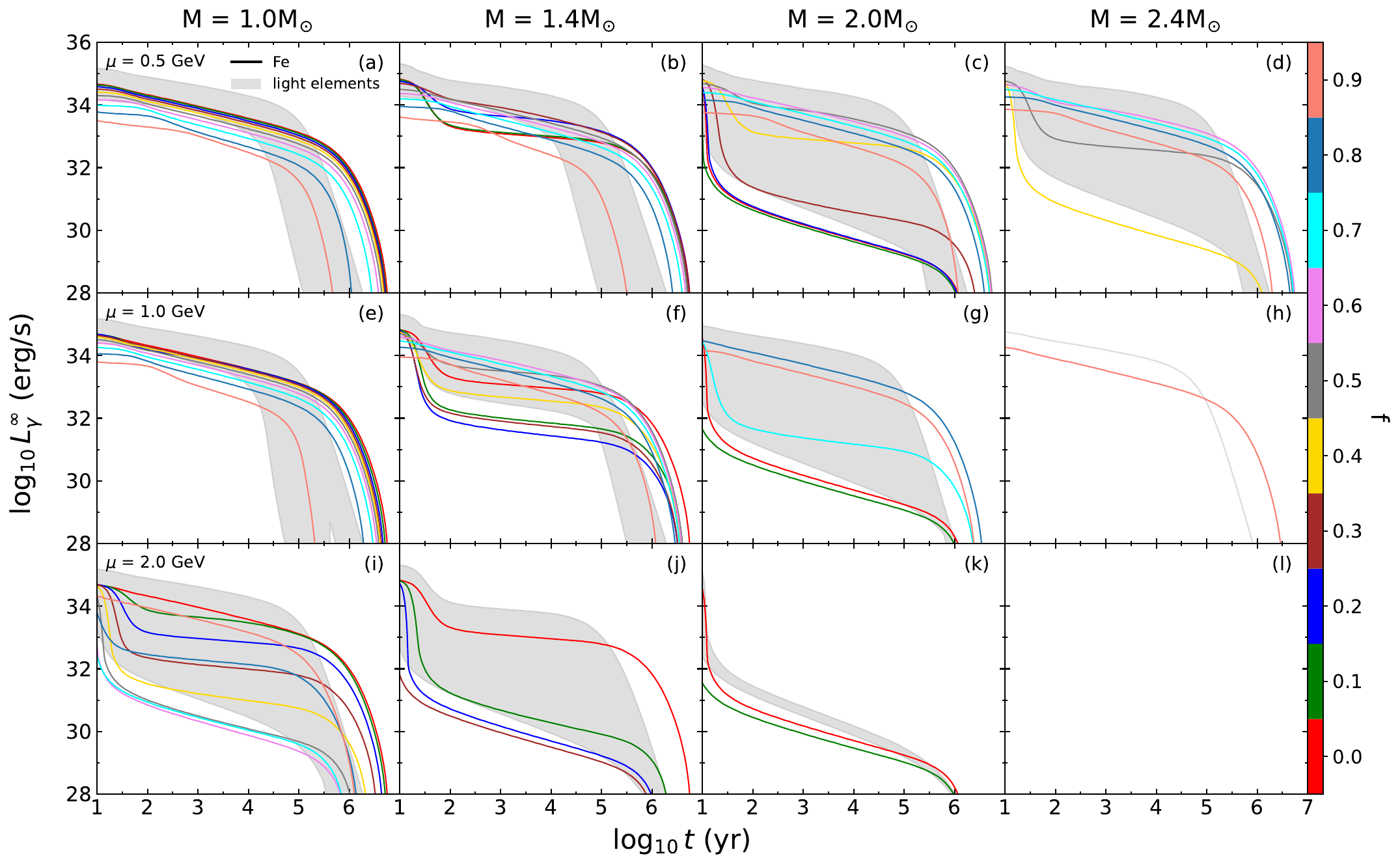}}
\vskip-4mm
\caption{
Cooling curves for DNSs with
different masses $M/\!\ms=1.0,1.4,2.0,2.4$ (columns)
and $\md=0.5,1,2\gev$ (rows),
varying $f=0,0.1,...,0.9$
in the permissible ranges according to Fig.~\ref{f:mrf}.
All results are obtained with $s_x=0.8,s_y=0.6$ p1S0 gap
and a Fe (curves) or light-elements (shading) atmosphere.
See text for details.
}
\label{f:coolf}
\end{figure*}

\begin{figure*}[t]
\vskip-3mm
\centerline{\hskip0mm\includegraphics[scale=0.58]{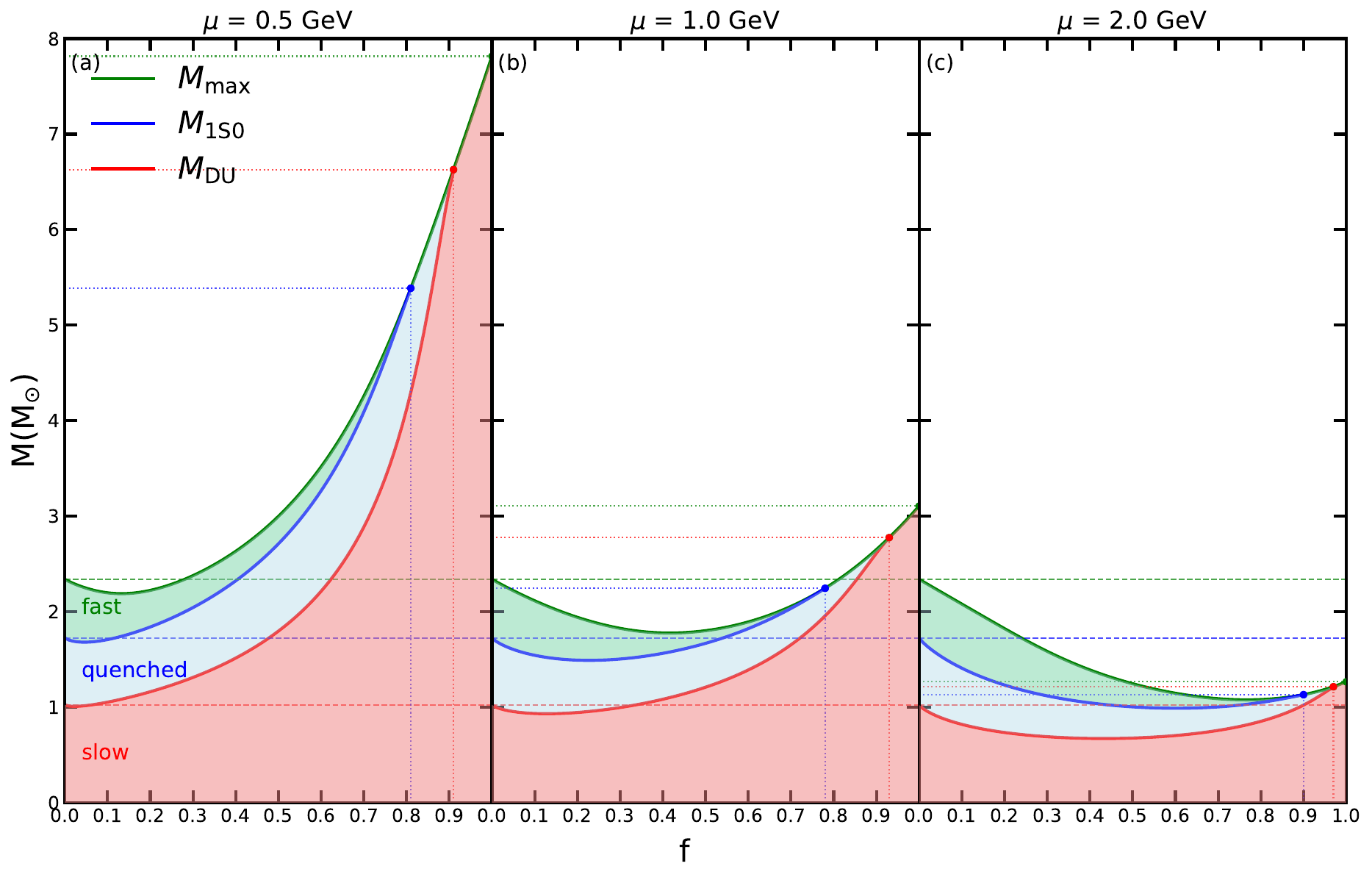}}
\vskip-5mm
\caption{
The maximum gravitational mass $\mmax$
and the threshold masses $\mdu$ and $\m1s0$
as functions of DM fraction $f$
for $\md=0.5,1,2\gev$.
Horizontal dashed lines indicate the values for pure NSs.
The regions of slow ($M<\mdu$),
quenched ($\mdu<M<\m1s0$),
and fast ($\m1s0<M<\mmax$) cooling are shown.
Their limits are indicated by vertical dotted lines.
\hj{}
}
\label{f:mduf}
\end{figure*}

\begin{figure*}[t]
\vskip-2mm
\centerline{\hskip0mm\includegraphics[scale=0.31]{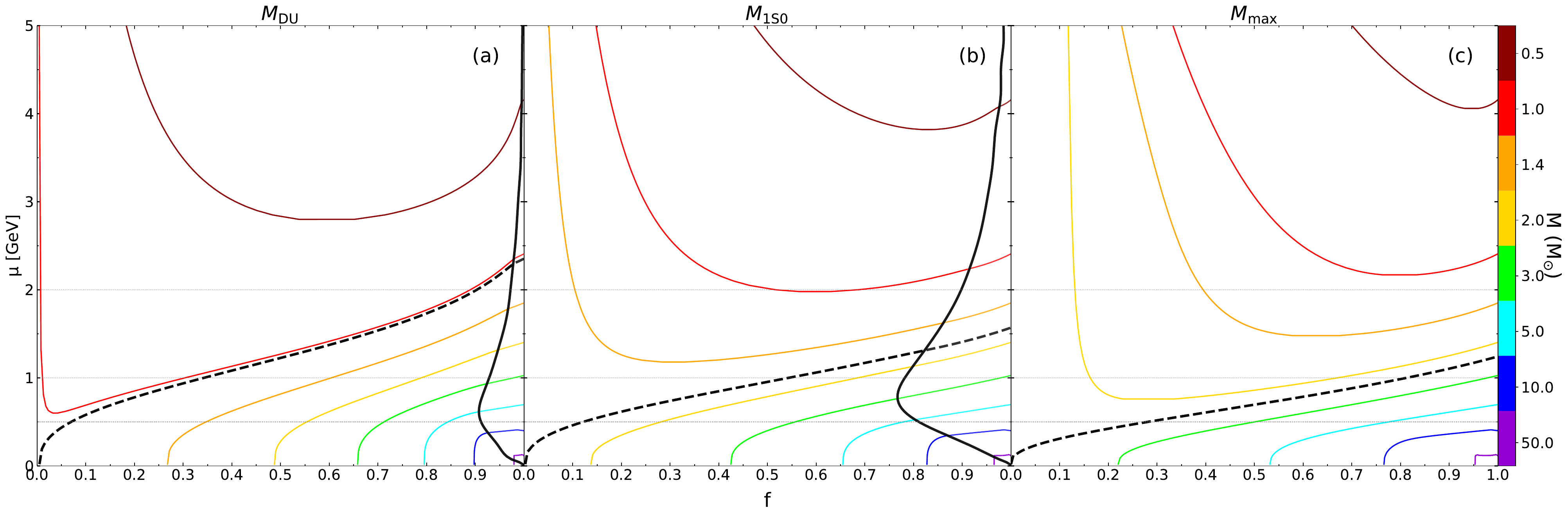}}
\vskip-4mm
\caption{
Contour plots of
$\mdu$ (upper limit of slow cooling),
$\m1s0$ (lower limit of fast cooling),
and $\mmax$
in the $(f,\md)$ plane.
The values $[\mdu,\m1s0,\mmax] = [1.03,1.73,2.34]\ms$
of a pure NS are indicated as dashed contours.
The solid black lines indicate the $\mdu,\m1s0=\mmax$ configurations,
respectively.
Horizontal lines at $\md=0.5,1,2$ facilitate comparison with Fig.~\ref{f:mduf}.
}
\label{f:mdufmu}
\end{figure*}


\section{Results}

\subsection{Equation of state and DNS structure}
\label{s:resa}

In order to illustrate the relevant properties of the chosen nuclear V18 EOS,
we show in Fig.~\ref{f:xp}
the DU onset condition (a),
the proton fraction (b),
the p1S0 gap (c),
and the pure NS mass (d)
as a function of the nucleon (central) density~$\rho$
of betastable and charge-neutral NS matter.
For this EOS the proton fraction reaches the DU threshold $\xdu=0.135$
at a density $\rdu \approx 0.371\fm3$.
The associated NS mass with that central density is $\mdu=1.03\ms$,
hence nearly all NSs modeled with the V18 EOS can potentially cool very fast.
However, DU cooling is quenched by the  $s_x=0.8$ scaled BCS p1S0 gap,
which vanishes at $\r1s0=0.536\fm3$,
corresponding to the central density of a $\m1s0=1.73\ms$ NS.
Therefore, $M<\m1s0$ stars are superfluid throughout,
whereas for $M>\m1s0$ there is a growing non-superfluid core region,
allowing them to cool very fast again.


The value of the NS maximum mass $\mmax=2.34\ms$ for the V18 EOS
is compatible with the current observational lower limits
\cite{Antoniadis13,Arzoumanian18,Fonseca21};
in particular the recent data $M=2.35\pm0.17\ms$ for PSR J0952-0607
\cite{Romani22}.
The radii $R_{1.4}=12.32\km$ and
$R_{2.0}=11.93\km$
are compatible with the combined GW170817+NICER analysis
\cite{Rutherford24},
$R_{1.4}=12.28^{+0.50}_{-0.76}\km$ 
and
$R_{2.0}=12.33^{+0.70}_{-1.34}\km$. 

\OFF{
The most massive cold NSs observed so far are
PSR~J1614-2230 ($M=1.908\pm0.016\ms$) \cite{Arzoumanian18},
PSR~J0348+0432 ($M=2.01\pm0.04\ms$) \cite{Antoniadis13}, and
PSR~J0740+6620  2.14-0.09+0.10 \cite{Cromartie20}, OLD!
PSR~J0740+6620 ($M=2.08\pm0.07\ms$) \cite{Fonseca21},
PSR~J0952-0607 ($M=2.35\pm0.17\ms$) \cite{Romani22}

PSR~J0740+6620 with measured radius
$R(2.08\pm0.07M_\odot) = 13.7^{+2.6}_{-1.5}\km$ \cite{Riley21}
$R(2.072^{+0.067}_{-0.066}\ms) = 12.39^{+1.30}_{-0.98}\km$ \cite{Miller21}
$R(2.073\pm0.069\ms) = 12.49-0.88+1.28 \cite{Salmi24} 11.6,12.5,13.8
12.76-1.02+1.49 or 12.92-1.13+2.09 \cite{Dittmann24}

We also mention the combined estimates of the mass and radius
of the isolated pulsar J0030+0451 observed recently by NICER,
$M=1.44^{+0.15}_{-0.14}\ms$ and $R=13.02^{+1.24}_{-1.06}\km$
\cite{Miller19,Riley19}
$M=1.36^{+0.15}_{-0.16}\ms$ and $R=12.71^{+1.14}_{-1.19}\km$ 11.5,12.7,13.8
\cite{Miller21,Riley21}
$M=1.40^{+0.13}_{-0.12}\ms$ and $R=11.71^{+0.88}_{-0.83}\km$ 10.9,11.7,12.6
$M=1.70^{+0.18}_{-0.19}\ms$ and $R=14.44^{+0.88}_{-1.05}\km$
\cite{Vinciguerra24}

and in particular the result of the combined GW170817+NICER analysis
\cite{Abbott17,Abbott18},
$R_{2.08}=12.35\pm0.75\km$ \cite{Miller21}, and
$R_{1.4}=
$12.45\pm0.65$ km \cite{Miller21}, 
$11.94^{+0.76}_{-0.87}$ km \cite{Pang21}, and
$12.33^{+0.76}_{-0.81}$ km or
$12.18^{+0.56}_{-0.79}$ km \cite{Raaijmakers21}
$12.28^{+0.50}_{-0.76}$ km \cite{Rutherford24}PP
$12.01^{+0.56}_{-0.75}$ km \cite{Rutherford24}CS
$R_{2.0}=12.33^{+0.70}_{-1.34}$ km \cite{Rutherford24}PP
$R_{2.0}=11.55^{+0.94}_{-1.09}$ km \cite{Rutherford24}CS
}

Fig.~\ref{f:mrd} illustrates the EOS for pure dark stars ($f=1$)
with different particle masses $\md=0.5,1,2\gev$.
It compares the mass-radius relations of dark stars with that
of the nucleonic NS.
It can be seen that for $\md\approx1\gev$,
of the same order as the nucleon mass,
also the stellar mass is of the same order as that of a pure NS,
whereas for $\md <(>) 1\gev$,
both mass and radius become increasingly larger (smaller) than those
of pure NSs.
This is important for the later discussion,
as in the limit of large $f$ these dark-star properties are approached.
However, while the total mass $M$ is observable by its gravitational effects,
the DM radius $R_D$ is not directly observable,
but only the optical nucleonic radius $R_N$ is \cite{Liu24}.
This will be discussed in the following.

The properties of DNSs with arbitrary DM fraction $f$
are illustrated in Fig.~\ref{f:mrf},
which shows the DNS mass-radius relations for fixed DM fractions
$\fd=0,0.1,...,1$
(colored curves)
and for different DM particle masses $\md=0.5,1,2\gev$.
The solid black curves indicate the sequence of maximum masses
with increasing $\fd$,
whereas the broken black curves connect the $R_N=R_D$ positions $\mcr$
on the fixed-$\fd$ curves.
The insets visualize both curves in the $(M,\fd)$ plane,
where DM-core and DM-halo domains are indicated.
One can easily realize that
`small' DM particle masses $\md\lesssim1\gev$ permit DM-halo stars
with associated increase of maximum mass and gravitational radius
already at very small values of $f$,
whereas `large' DM particle masses $\md\gtrsim1\gev$
require large DM fractions to form DM-halo stars,
and their maximum masses and radii are always reduced
relative to the pure NS \cite{Liu24}.

Regarding the stability of the DNS configurations shown in Fig.~\ref{f:mrf}
against radial oscillation perturbations,
we show in Fig.~\ref{f:osc} the radial oscillation frequency
computed in the two-fluid framework \cite{Kain20,Kain21}
as a function of the DNS gravitational mass.
In all cases the frequency remains positive up to the maximum mass
to a very good numerical accuracy.
Further exotic stable configurations as studied in some recent works
\cite{Hippert23,Barbat24,Zhen24,Pitz25,Biesdorf25} are not excluded,
but disregarded in the present work dedicated to the cooling properties of DNSs.

For the cooling simulations the stellar radial profiles
of energy densities $\eps_N, \eps_D$
and nucleon density $\rho$ are of foremost importance,
and for fixed DNS gravitational mass $M=M_N+M_D$
they change with the DM parameters $\md$ and $f$.
In Fig.~\ref{f:nrm} we display some representative configurations
for fixed $f=0.5$ and $\md=0.5,1,2\gev$
over the possible range of $M$ according to Fig.~\ref{f:mrf}.
As evinced in that figure,
for $f=0.5$ all DNSs are DM-halo stars for $\md=0.5,1\gev$
and DM-core stars for $\md=2\gev$,
but the mass range is restricted to $M<1.25\ms$ in the latter case.
One can see a systematic reduction of the visible radii $R_N$
with increasing $\md$, i.e.,
for fixed $M$ (and $f$) the nucleonic core becomes more compact
and its average nucleon density increases.
This is due to the fact that with increasing $\md$ also the DM halo
shrinks and compactifies the enclosed nuclear matter more efficiently.
In particular the increase of the nuclear density is important for the cooling,
as for example the DU process becomes active at a smaller mass~$M$.
To see this,
the densities of DU onset $\rdu$ and of vanishing p1S0 gap $\r1s0$
are indicated in the figure.
DNSs with a nuclear density profile remaining below $\rdu$ cool very slowly,
reaching between $\rdu$ and $\r1s0$ their DU cooling is quenched by pairing,
and above $\rdu$ they are cooling fast by the unblocked DU process.

This is more clearly seen in
Fig.~\ref{f:nrf} that shows the same information for
a fixed $M=1.4\ms$ and varying $f=0,0.1,\ldots,1$.
For example, for $\md=1\gev$ an increase of $f$ (up to about 0.4)
leads to an enhancement of the central nucleon density
by reduction of the nuclear radius,
counteracting the reduction of the total nuclear matter in the star.
Only for larger $f$ the latter dominates.
In this particular case, unblocked DU cooling is still not reached, though,
but the cooling becomes very slow for $f\gtrsim0.6$.
These effects depends strongly on the DM particle mass $\md$,
as discussed before.
In fact for $\md=2\gev$ the $M=1.4\ms$ DNSs with $f>0.104$ cool very fast,
but only values of $f<0.395$ are possible in this case,
see Fig.~\ref{f:mrf}.


These examples demonstrate the rich variety of DNSs
with degrees of freedom $\md$ and $f$,
and their cooling properties.
We will provide a more general analysis later.

\subsection{Cooling simulations}
\label{s:resb}

We now present the results of cooling simulations for DNSs that
contain an arbitrary fixed DM fraction $f$.
As stressed before,
such analysis must be based on a nuclear EOS and microphysics
that is able to reproduce in a reasonable way
{\em all} currently known cooling data of pure NSs.

We have developed such a framework in \cite{Wei20,Das24},
where we have shown that
an EOS featuring DU cooling for a wide enough mass range of NSs
together with (partial) quenching by the p1S0 BCS gap
seems to be required to reproduce the current cooling data.
It seems difficult to accommodate finite n3P2 pairing in this setup.
More precisely, we use the V18 EOS including p1S0 pairing
with optimized scaling parameters
$s_x=0.8,s_y=0.6$, Eq.~(\ref{e:sxy}),
that we found in \cite{Wei20,Das24} to give best results
for deduced NS mass distributions.
No n3P2 pairing is included,
since the associated n3P2 PBF process provides a too strong cooling
for old ($\gtrsim10^6$yr) objects \cite{Wei20b},
with or without DU process,
as already found in several previous works
\cite{Grigorian05,Beznogov15,Beznogov15b,Beznogov18,Potekhin19,Potekhin20,Wei20}.
We extend now this setup by adding DM.


Fig.~\ref{f:coolm} shows the cooling diagrams
obtained for DNSs with DM fractions $f=0,0.1,0.5,0.9$ (columns),
employing a Fe atmosphere (solid curves)
or a light-elements atmosphere
(spanning the shaded background areas),
in comparison with the currently known data points
\cite{Beznogov15,Potekhin20,Cooldat}
comprising partially large (estimated) error bars,
shown only in the $f=0$ panels
for comparison with the pure NS results.
The upper row displays the results obtained without superfluidity
and the lower row includes optimal $s_x=0.8,s_y=0.6$ p1S0 pairing.
\OFF{
namely for
weakly-magnetized NSs (WM, green),
ordinary pulsars (OP, blue),
high-magnetic-field-B pulsars (HB, pink),
the magnificent seven (MS, violet),
small hot spots (HS, magenta), and
upper limits (UL, orange).
There are altogether 57 data points,
}

Summarizing briefly the results for pure NSs (left column, $f=0$),
as obtained in previous works
\cite{Taranto16,Fortin18,Wei19,Wei20,Das24},
the results without pairing are clearly unrealistic,
as only the $M=1\ms$ stars cool slowly,
and all others too fast due to the early DU onset
at $\mdu=1.03\ms$ for the V18 EOS.
Inclusion of p1S0 pairing yields very reasonable cooling curves
that cover all data points
(taking into account the uncertainty of atmospheres)
by quenching the DU process in the range
$1.03\ms = \mdu < M < \m1s0 = 1.73\ms$,
while only high-mass stars, $M>1.73\ms$, still cool very rapidly.
In fact the $s_x=0.8,s_y=0.6$ p1S0 scaling parameters were optimized
to give the best agreement with current models of NS mass distributions,
see \cite{Wei20,Das24} for details.

Based on these results the understanding of the effect of DM
(panels $f=0.1,0.5,0.9$)
is straightforward:
The results without pairing are still dominated by the contrast
between heavy (light) stars with(out) DU cooling,
with a sharp threshold.
However, the threshold mass depends now on the DM fraction, $\mdu(f)$,
shown in Fig.~\ref{f:mduf}(b):
For large DM fraction, the DU onset is shifted to large
gravitational masses due to the dilution of nuclear matter inside the star.
At the same time the maximum mass of the DNS increases at large $f$,
as seen in Fig.~\ref{f:mrf}.
For example, for $f=0.5,0.9$ the threshold becomes $\mdu=1.210,2.613\ms$,
which is clearly seen in the relevant panels (c) and (d) of Fig.~\ref{f:coolm},
where only the $M\geq1.3,2.7\ms$ stars cool fast, respectively.
In the realistic case with pairing (lower row),
the sequence of the cooling curves remains continuous,
but now also $\m1s0$ depends on $f$
and its increase towards $\mmax$ at large $f$
restricts the possible range of fast DU cooling of massive stars.
Indeed Fig.~\ref{f:mduf}(b) shows that for $\md=1\gev$
fast cooling is only possible up to $f=0.78$,
whereas at $f=0.9$ only slow or quenched cooling is possible,
as confirmed in Fig.~\ref{f:coolm}(h).


It is more interesting to study the cooling properties of DNSs
with fixed mass $M$ and varying $f$. 
Such results are shown in Fig.~\ref{f:coolf}
for fixed masses $M=1.0,1.4,2.0,2.4\ms$.
While Fig.~\ref{f:coolm} showed only $\md=1\gev$ results,
we add now those for $\md=0.5,2\gev$,
which adds complexity to the cooling scenarios.
Only realistic cooling curves including pairing
are shown.

Beginning again with the $\md=1\gev$ results (2nd row),
and referring to the cooling domains in Fig.~\ref{f:mduf}(b),
we note that\\
-- $M=1\ms$ (e):
all stars lie in the slow-cooling (or strongly quenched)
domain for any~$f$.
Nevertheless the cooling effect increases with increasing $f$,
since the cooling nucleonic mass decreases;
for $f=0.1$ it is only $M_N=0.1\ms$.\\
-- $M=1.4\ms$ (f):
there is a transition from quenched cooling to slow cooling
at $f=0.604$, clearly seen in the panel.\\
-- $M=2\ms$ (g):
such DNSs exist only for $f<0.163$ and $f>0.671$,
see Fig.~\ref{f:mduf}.
The former are fast cooling, the latter quenched or slow cooling.\\
-- $M=2.4\ms$ (h):
only exists for $f>0.831$.
Slow cooling for $f>0.864$.

The properties of the $\md=0.5,2\gev$ DNSs are more extreme:
Fig.~\ref{f:mduf} shows that for $\md=2\gev$ all three characteristic masses
decrease over a large range of $f$;
therefore the cooling properties correspond to those
of normal NSs with scaled-down mass.
On the contrary,
for $\md=0.5\gev$ the characteristic masses increase with $f$,
and the cooling corresponds to scaled-up NSs.
Those features can be verified comparing Figs.~\ref{f:coolf}
and \ref{f:mduf}.
Some interesting extreme cases are\\
-- $M=2.4\ms$ (d,h,l):
This star does not exist as pure NS.
For $\md=0.5\gev$, it is possible for $f>0.311$
and for $f>0.633$ is slowly cooling, in spite of its large mass.
For $\md=1\gev$, only the slowly cooling $f=0.9$ configuration exists,
and for $\md=2\gev$, $\mmax(f)<\mmax(0)<2.4\ms$.\\
--  $M=1\ms$ (a,e,i):
It is slowly cooling for $\md=0.5,1\gev$,
but fast cooling for $\md=2\gev$ and $0.517<f<0.686$,
in spite of its small mass.


We have thus shown that due to the two free parameters $\md$ and $f$
a very rich scenario of DNSs and their cooling properties exists.
In order to draw some general conclusions,
we finally show in Fig.~\ref{f:mdufmu} the contour plots of
$\mdu$, $\m1s0$, and $\mmax$ in the $(f,\md)$ plane.
The first indicates the upper limit for slow cooling without DU process,
the second the upper limit for quenched DU cooling
and thus the lower limit for fast unblocked DU cooling.
DNSs with $M<\mdu$ cool slowly and those with $M>\m1s0$ cool fast.

The results generalize those in Fig.~\ref{f:mduf};
for example, taking the yellow $\mdu=2\ms$ contour in panel 9(a),
all DNS $(f,\md)$ configurations with that mass
lying below that curve cool slowly,
in spite of the large mass.
The solid black boundary line indicates the threshold $\mdu=\mmax$,
where the $\mdu$ contour transits onto the same $\mmax$ contour
shown in panel 9(c),
see also Fig.~\ref{f:mduf}.
All configurations on the right side of that line cool slowly.
The dashed black curve indicates the $\mdu=1.03\ms$ contour
corresponding to the value of the pure NS.
Therefore the $(f,\md)$ domain below that line permits slow cooling of DNSs
heavier than a pure NS.

Similarly, all DNS $(f,\md)$ configurations with a mass $M=1\ms$
lying above the red $M=1\ms$ contour in panel 9(b) cool fast,
in spite of the small mass.
All configurations to the right of the solid black $\m1s0=\mmax$ boundary
cool quenched by the p1S0 gap.
All $(f,\md)$ DNSs
above the dashed black $\m1s0=1.73\ms$ contour
permit fast cooling of DNSs lighter than a pure NS.

Summarizing, the domain below the dashed black line in (a)
allows slow cooling for `atypically' heavy stars,
while the domain above the dashed black line in (b)
allows fast cooling for `atypically' light stars.
We stress again that the $f>0$ predictions should not be compared
with the current set of cooling data,
but would be appropriate for individual rare exotic DNSs
with large DM fraction
that have probably not been observed so far.

\section{Conclusions}
\label{s:end}

We have investigated the cooling properties of hypothetical DNSs
with large DM fractions,
up to pure dark stars.
We employed a standard DM EOS of self-interacting sterile fermionic DM
satisfying an observational constraint on the self-interaction cross section,
and thus characterized by only one parameter, the DM particle mass $\md$.
The nucleonic BHF V18 EOS features a very low DU onset density,
and cooling has to be moderated by p1S0 pairing
with optimized scaling factors,
such that the current set of cooling data is reasonably well explained.

We computed cooling curves for DNSs with varying DM fraction $f$
and DM particle mass $\md$.
In general fast cooling occurs beyond the mass $\mdu$
corresponding to the DU onset density $\rdu$
of the nuclear matter inside the star,
but is blocked by pairing up to $\m1s0$
corresponding to the p1S0 gap vanishing at the central density $\r1s0$.
These quantities depend on the DM parameters $f$ and $\md$
and thus predominantly control the cooling.
At the same time, also the maximum DNS gravitational mass
depends on the DM parameters,
and the interplay of these features
determines the rich variety of cooling DNSs.

One can roughly discriminate two cases depending on the DM particle mass:\\
(a) `Large' $\md>1\gev$:
mostly DM-core stars,
small $\mmax(f)<\mmax(0)$,
but also small $R_N(f)<R_N(0)$
such that the nucleonic density remains sufficiently large to
permit fast cooling in many cases.
The control parameters $\mdu,\m1s0,\mmax$
generally decrease with DM fraction $f$,
therefore these are low-mass DNSs
with a full spectrum of cooling possibilities.
In particular, there are fast-cooling very light stars,
in contrast to ordinary NSs.\\
(b) `Small' $\md<1\gev$:
mostly DM-halo stars,
large $\mmax(f)>\mmax(0)$,
small change of nucleonic radii, $R_N(f) \approx R_N(0)$,
therefore nucleonic density and cooling strength
depend sensibly on both $M$ and $f$.
$\mdu,\m1s0,\mmax$ generally increase with DM fraction $f$,
therefore these are `blown-up' cooling NSs
with the unique possibility of slow-cooling but very heavy stars,
due to their large DM content.

In many cases it is thus not possible to distinguish
the cooling properties of a DNS with unknown $\md$ and $f$
from those of a pure NS of the same mass.
However, if a very massive
(possibly much more than the NS $\mmax$)
and slow-cooling object would be observed,
it could be an indication for a DNS with small $\md$ and large $f$
(and therefore low nuclear density).
On the contrary, a very light but fast-cooling star would indicate
large $\md$ and not too large $f$,
such that $\m1s0(f)$ is exceeded before reaching the $\mmax(f)$ limit.

The results presented in this work are specific for the V18 nuclear EOS,
regulated by the parameters
$[\mdu, \m1s0, \mmax] = [1.03,1.73,2.34]\ms$,
but the general qualitative conclusions expressed above
remain the same for any nuclear model
comprising DU cooling and pairing
with different parameter values.


The DM model in this work was fixed to a fermionic one with self interaction.
However, currently many other choices are possible,
in particular those involving interaction with ordinary matter,
and therefore providing thermal coupling with the nuclear matter
and additional direct cooling processes.
This would still enhance the complexity
of what has been discussed in this work.

Also, exotic types of ordinary matter like hyperons or quark matter
were not considered in this work.
Those still present a formidable challenge to cooling calculations
due to their largely unknown microphysics ingredients relevant for cooling,
like cooling rates, heat capacities, transport properties,
and most of all superfluid properties.

NS cooling remains a fascinating but very complex and
difficult to control topic.
Future new data,
in particular on masses of the cooling objects,
will however severely tighten the constraints on the theoretical models.

\vspace{10mm}

\acknowledgments{
This work was partially funded by the
National Key R\&D Program of China No.~2022YFA1602303 and the
National Natural Science Foundation of China under
Grants Nos.~12205260, 12147101, and 11975077.
}


\newcommand{\araa}{Annu. Rev. Astron. Astrophys.}
\newcommand{\aap}{Astron. Astrophys.}
\newcommand{\apjl}{Astrophys. J. Lett.}
\def\epja{Eur. Phys. J. A}
\def\epjc{Eur. Phys. J. C}
\def\jpg{J. Phys. G}
\newcommand{\mnras}{Mon. Not. R. Astron. Soc.}
\def\npa{Nucl. Phys. A}
\newcommand{\physrep}{Phys. Rep.}
\newcommand{\plb}{Phys. Lett. B}
\def\ppnp{Prog. Part. Nucl. Phys.}
\newcommand{\ssr}{Space Science Reviews}

\bibliography{cooldm} 

\end{document}